\documentclass{article}

\usepackage{amsmath}

\usepackage{amssymb}

\usepackage{cite}

\title{On finite time singularities in scalar field dark energy models based in the RS-II Braneworld}

\author{Oem Trivedi$^a$ \footnote{oem.t@ahduni.edu.in}  \and Maxim Khlopov \footnote{khlopov@apc.in2p3.fr}  $^{b,c,d}$}
\date{%
	$^a$School of Arts and Sciences, Ahmedabad University,Ahmedabad 380009,India\\%
	$^b$Institute of Physics, Southern Federal University,
	Stachki 194 Rostov on Don 344090, Russia\\%
	$^c$Universite de Paris, CNRS, Astroparticule et Cosmologie
	F-75013 Paris, France\\%
	$^d$National Research Nuclear University ”MEPHI” 115409 Moscow, Russia\\%
	\today
}

\begin{document}

	\maketitle
	
	\section{Introduction}
	
	Observations of late time acceleration of the Universe took surprised the whole cosmological community \cite{SupernovaSearchTeam:1998fmf} . Ever since then a lot of work has been done in order to explain this expansion which include the standard ones like the Cosmological constant \cite{SupernovaSearchTeam:1998fmf,Weinberg:1988cp,Lombriser:2019jia,Copeland:2006wr,Padmanabhan:2002ji} alongside more exotic scenarios like Modified gravity theories\cite{Capozziello:2011et,Nojiri:2010wj,Nojiri:2017ncd}and very appealing ways of detecting dark energy directly have been put forward recently as well \cite{Zhang:2021ygh}. An exciting approach towards understanding dark energy is that of Quintessence, where a scalar field drives the late-time cosmic acceleration of the universe\cite{Zlatev:1998tr,Tsujikawa:2013fta,Faraoni:2000wk,Gasperini:2001pc,Capozziello:2003tk,Capozziello:2002rd,Carroll:1998zi,Caldwell:2005tm,Han:2018yrk,Astashenok:2012kb,Shahalam:2015sja}. Quintessence is very interesting in the sense that it is the simplest scalar field dark energy scenario which is not plagued with problems like ghosts or Laplacian instabilities. In quintessence models a slowly varying scalar field with some potential $V(\phi)$ leads to the acceleration of the universe, with the mechanism being similar to slow roll inflation with the difference being that in this case we cannot ignore contributions from non-relativistic matter like Baryons and Dark Matter. This should be noted too, however, that simple models of Quintessece have been shown to be at odds with the current H0 tension \cite{Colgain:2019joh,Banerjee:2020xcn,DiValentino:2021izs} and hence simple models of Quintessence seem to perform worses than Lambda-CDM models with regards to the current H0 data\cite{riess2021comprehensive}.  
	\\
	\\
	Quintessence has been studied in a variety of exotic cosmological scenarios as well like in the context of various modified gravity theories alongside cosmological models which are modified by quantum gravity based corrections like the RS-II Braneworld. Braneworld cosmologies is especially interesting because it is a class of theories that confines the particles of the Standard Model to a (3+1) dimensional brane embedded in
	a higher-dimensional bulk with compactified dimensions. Interestingly, gravity is not confined to the brane and can propagate through the extra dimensions and this feature provides a novelty to brane theories from other extra dimensional theories of gravity. The RS-II model is a based on a modification of the RS-I Braneworld cosmology model\cite{Randall:1999ee}, where the hierarchy problem is solved by embedding two 3-branes in a five-dimensional bulk where one of the branes contains the Standard Model particles. The RS-II braneworld cosmology removes one of the 3-branes and recovers both Newtonian gravity and General Relativity as its limiting cases\cite{Randall:1999vf}. Since another braneworld cosmology scenario in the form of Dvali-Gabadadze-Porrati (DGP)  model \cite{Deffayet:2000uy,Dvali:2000hr} can produce some effects on the late-time evolution of the universe, it becomes very interesting to see if one can address late-time acceleration issues using the RS-II model as well. In this direction, there have been several works in recent times which have addressed  Quintessence in the RS-II Braneworld scenario\cite{Sahni:2002dx,Sami:2004xk,Bento:2008yx}. There have been a lot of papers which have tried a dynamical systems approach analysis of these models as well. For example, in \cite{Gonzalez:2008wa,Leyva:2009zz} such an analysis was performed with the inclusion of a scalar field confined to the brane and it was then shown that the canonical scalar only affects the early-time behaviour of the universe, and that in inflationary critical points exist for a constant scalar potential. While in \cite{Escobar:2011cz} centre manifold theory was used with a wide variety of potentials to study the asymptotic behaviour of RS-II models.
	\\
	\\
	There has also been an expansive literature in recent times which has been devoted study the various types of singularities that could occur during the current and far future of the Universe \cite{Nojiri:2004ip,Nojiri:2005sr,Nojiri:2005sx,Bamba:2008ut,Bamba:2010wfw,Nojiri:2008fk,odintsov2022did}. Talking from a dynamical point of view, one of the most interesting aspects of various dynamical systems that one can investigate in them is their singularity structure which is all the more relevant when the dynamical systems describe physically interesting phenomena.  While there has been a lot of work which has discussed methods to explore the singularity structure of autonomous dynamical systems, a particularly interesting one is the procedure of Goriely and Hyde \cite{goriely2000necessary}. As cosmology puts forward a lot of very intriguing dynamical systems, the exploration of the singularity structure of such systems have garnered a lot of interest in recent times and this analysis method has been particularly useful for such explorations in cosmology\cite{barrow2004more,cotsakis2007dominant,cotsakis2007asymptotics,antoniadis2010brane,antoniadis2013brane,antoniadis2014enveloping}. This has previously been applied to study finite time singularities in certain classes of quintessence models too\cite{odintsov2018dynamical} and \cite{odintsov2019finite} also discussed the implications of the Swampland dS conjecture on the singularity structure. The Swampland refers to the class of low energy EFT's which would not have a consistent UV completion with regards to String theory, and hence potentially with quantum gravity as a whole \cite{vafa2005string}. In order to classify which theories should belong to the swampland, a number of swampland conjectures have came forward in recent years. One of these conjectures is the dS conjecture \cite{Ooguri:2016pdq}, which is based on the rather bold idea that no meta stable dS spaces can be found in the landscape of string theory or quantum gravity as a whole. It can be written in Planck Units $m_{p} = 1 $, as \begin{equation}
		\frac{|V^{\prime} (\phi)|}{V(\phi)} \geq \mathcal{O} (1)
	\end{equation} where we V($\phi$) is the potential of the scalar field theory. This has had some very interesting implications for dark energy too \cite{Agrawal:2018own,Chiang:2018lqx,Trivedi:2020xlh}. So in this paper we will apply the Goriely-Hyde singularity analysis method on Quintessence models based in a RS-II Braneworld cosmology and explore the singularity structure of these models and would try to see if the swampland dS conjecture has any significant effect on the singularity structure as well. We will consider a scalar field described by a general action, as considered in \cite{Dutta:2015jaq,Ravanpak:2020ewi} which allow us to describe quintessence and phantom fields at the same time. In Section II we will briefly describe the Goriely-Hyde procedure after which we will discuss in the detail about the occurrence of finite time singularities in RS-II Braneworld scalar field dark energy models through the dynamical systems view. After this, we will consider a well motivated ansatz for the Hubble parameter and show that these regimes of dark energy can allow for weak singularities of the Type III and Type IV class and can also allow for strong singularities like the Big Rip (Type I). We will finally conclude our work in section IV.
	\\
	\\
	\section{Goreily-Hyde method}
	The Goriely-Hyde singularitiy analysis method \cite{goriely2000necessary} is a very elegant way to ascertain the existence of finite-time singularities in dynamical systems. We can describe the procedure in a step-wise way as follows : \begin{itemize}
		\item We start by considering a dynamical system of n differential equations of the form, \begin{equation}
			\dot{x}_{i} = f_{i}(x)
		\end{equation}
		where $ i =1,2,.n $ and the overdot denotes a differentiation with respect to time t which in the case of quintessence models would be better represented by the number of e-foldings N. We can then extract from $f_{i} $ parts of the equation that become more significant as one reaches the region of the singularity (such parts are called as the "dominant parts" \cite{goriely2000necessary}). Each dominant part constitutes a mathematically consistent truncation of the system and we can denote these parts as $\hat{f}_{i}$. Now the system takes the form \begin{equation}
			\dot{x}_{i} = \hat{f}_{i}(x)
		\end{equation} 
		\item Without any loss of generality, the $x_{i}'s$ near the singularity will assume the form \begin{equation}
			x_{i} = a_{i} \tau^{p_{i}}
		\end{equation}
		where $\tau = t - t_{c} $ with $t_{c} $ being an integration constant. We can then substitute (4) in (3) and equate the exponents to find values of $p_{i}$ for various i which will constitute the vector $ \mathbf{p} = (p_{1},p_{2},...p_{n}) $ after which we can calculate all the values of $ a_{i} $ similarly to have $ \vec{a} = (a_{1},a_{2},...a_{n} ) $. Note that if $ \vec{a} $ has only real entries then it will give rise to only finite-time singularities while if it contains even one complex entry, then it may give rise to only non finite-time singularities. Taking this into account, every set ($a_{i},p_{i} $) is known as a dominant balance of the system. 
		\item After this, one calculates the Kovalevskaya matrix, which is given by
		\begin{equation}
			R = \begin{pmatrix}
				\frac{\partial f_{1}}{\partial x_{1}} & \frac{\partial f_{1}}{\partial x_{2}} & . & . & \frac{\partial f_{1}}{\partial x_{n}}\\
				\frac{\partial f_{2}}{\partial x_{1}} & \frac{\partial f_{2}}{\partial x_{2}} & . & . & \frac{\partial f_{2}}{\partial x_{n}}\\
				. & . & . & . & . \\
				. & . & . & . & . \\
				\frac{\partial f_{n}}{\partial x_{1}} & \frac{\partial f_{n}}{\partial x_{2}} & . & . & \frac{\partial f_{n}}{\partial x_{n}}\\
			\end{pmatrix} -  \begin{pmatrix}
				p_{1} & 0 & . & . & 0 \\
				0 & p_{2} & . & . & 0 \\
				. & . & . & . & . \\
				. & . & . & . & . \\
				0 & 0 & . & . & p_{n} \\
			\end{pmatrix}
		\end{equation} 	
		After finding this matrix, one has to evaluate this in different dominant balances and find their eigenvalues. The eigenvalues would then have to be of the form $(-1,r_{2},r{3},..,r_{n}) $ and $ r_{2},r_{3}..>0$ then the singularity is general and will occur irrespective of the initial conditions of the system. On the other hand, even if one of $ ( r_{2},r_{3}..)$ are negative, then the singularity is local and will only occur for certain sets of initial conditions.
	\end{itemize} 
	\section{Singularity analysis in RS-II scalar field models}
	The scalar that we will consider here is described by a general action, as considered in \cite{Dutta:2015jaq,Ravanpak:2020ewi} which will allow us to describe quintessence and phantom fields at the same time. We can write the total action of the RS-II Model inclusive of both the scalar and the background fluid term as \begin{multline}
		S  = S_{RS} + S_{B} + S_{\phi} = \int d^5 x \sqrt{-g^{(5)}} \left( \Lambda^{(5)}  + 2 R^{(5)}   \right) + \\ \int d^4 x \sqrt{-g} \left(\sigma -\frac{1}{2} \mu(\phi) (\nabla \phi)^2 - V(\phi)  + \mathcal{L}_{B}  \right)
	\end{multline}
	where $R^{(5)} $, $ g^{(5)}_{\mu \nu} $ and $ \Lambda^{(5)} $ are the bulk Ricci Scalar, metric and the cosmological constant respectively with $\sigma$ being the brane tension on the 3-brane, $g_{\mu \nu} $ being the 3-brane metric and $\mu(\phi) $ being a scalar coupling function. Note that here we are working in Planck units with $(m_{p}^{(5)})^2 = 1 $ with $ m_{p}^{(5)} $ being the 5-dimensional Planck mass. Assuming that the brane metric has the usual FLRW form, we get the Friedmann equation to be \cite{Maartens:2010ar} \begin{equation}
		H^{2}  = \rho \left(1 + \frac{\rho}{2 \sigma}\right)
	\end{equation} 
	where $ \rho = \rho_{\phi} + \rho_{B} $ is the total cosmological energy density taking into account contributions from both the scalar field and the background fluid term and the Bulk cosmological constant has been set to zero for simplicity. While pursuing the singularity structure of RS-II models with non-zero cosmological constant would certainly be an appealing endeavour, we are not interested in pursuing that here as we are interested in seeing how singularity structures of reasonably simple dark energy models in RS-II scenario plays out. By doing so we  are not imparting extreme fine-tuning to the models that could qualify for the forthcoming analysis, as it will be evident later on that the analysis we will pursue applies to a large range of Brane tensions , potentials and even for both quintessence and phantom forms of dark energy evolution. The second Friedmann equation can be written as \begin{equation}
		2 \dot{H} = -\left(1 + \frac{\rho}{\sigma}\right) \left(\mu(\phi) \dot{\phi}^2 + \rho_{B} \right)
	\end{equation}
	And the equation motion of the scalar is given by \begin{equation}
		\mu(\phi) \ddot{\phi} + \frac{1}{2} \frac{d \mu}{d \phi} \dot{\phi}^2  + 3 H \mu(\phi) \dot{\phi} + \frac{dV}{d\phi} = 0 
	\end{equation}
	Finally, using the following variables introduced in \cite{Gonzalez:2008wa} \begin{equation}
		x = \frac{\dot{\phi}}{\sqrt{6}H} \qquad y = \frac{\sqrt{V}}{\sqrt{3} H} \qquad z = \frac{\rho}{3 H^2}
	\end{equation}
	Choosing the background fluid to be of the form of pressurelees dark matter, in a way that $w_{B} = 0 $ \footnote{While this choice might appear ad-hoc on first sight, braneworld models with pressureless dark matter have been vividly discussed in recent literature \cite{Bouhmadi-Lopez:2002vho,Neves:2003eu,Kanno:2007wj,Banerjee:2011wk,Rani:2021hvh}. Besides this, the reason we are considering a dust-like form of the bulk matter is for the same reason as to why we didn't consider a non-zero bulk cosmological constant and that is for the sake of simplicity as we are interested in the singularity structure of reasonably simple dark energy models in the RS-II scenario. Singularity endeavours with bulk matter having considerable pressure would indeed make for a very nice exploration but we are not interested in pursuing that here in the sense that we want to see how vivid the singularity structure of models with simple form of bulk matter would look like. } , we get the dynamical system for this model to be \begin{equation}
		x^{\prime} = - \sqrt{\frac{3}{2 \mu }} \lambda y^2 -3x + \frac{3x}{2} \left(z + x^2 -y^2 \right) \left(\frac{2}{z} - 1 \right)
	\end{equation}
	\begin{equation}
		y^{\prime} = \sqrt{\frac{3}{2 \mu }} \lambda xy + \frac{3y}{2} \left(z + x^2 -y^2 \right) \left(\frac{2}{z} - 1 \right)
	\end{equation}
	\begin{equation}
		z^{\prime} = 3 (1-z) (z + x^2 - y^2 )
	\end{equation}
	where the primes denote differentiation with respect to the e-folding number N and $\lambda = \frac{V^{\prime}}{V}$, which in view of the dS conjecture would be constrained as $ \lambda \geq \mathcal{O}(1)$. Now that we have a proper autonomous dynamical system, we can start with the analysis. 
	The first truncation that we consider is given by \begin{equation}
		\hat{f} = \begin{pmatrix}
			-k \lambda y^2 \\
			-3 y^3 z^{-1} \\
			3 x^2
		\end{pmatrix}
	\end{equation}
	where $ k = \sqrt{\frac{3}{2 \mu}} $. Using the ansatz of the Goriely-Hyde method, we get $ \mathbf{p} = (-1,-1,-1) $ and using these, we get \begin{equation}
		\begin{aligned}
			a_{1} = \left(- \frac{1}{k \lambda} , \frac{i}{k \lambda} , - \frac{3}{k^2 \lambda^2} \right) \\[10pt]
			a_{2} = \left( - \frac{1}{k \lambda} ,- \frac{i}{k \lambda} , - \frac{3}{k^2 \lambda^2} \right)  
		\end{aligned}
	\end{equation}
	as both $ a_{1} $ and $a_{2}$ have complex entries, only non-finite time singularities will be possible with regards to this truncation. The Kovalevskaya matrix then takes the form \begin{equation}
		R =  \begin{pmatrix}
			1 & -2k\lambda y & 0 \\
			0 & 1 - \frac{9 y^2}{z} & \frac{3 y^{3}}{z^2} \\
			6x & 0 & 1 
		\end{pmatrix}
	\end{equation}
	We then finally find the eigenvalues of the matrix, which are given by \begin{equation}
		r = (-1,-1,2)
	\end{equation}
	Hence the singularities in this case will only be local singularities which will only form for a limited set of initial conditions. Note that we have not set any constraint on the value of $\lambda$ and so it does not matter what value $\lambda$ takes in this case. Hence this result will hold for values of $\lambda$ which are both swampland consistent and swampland inconsistent with regards to the dS conjecture(1).
	\\
	\\
	The second truncation that we consider is given as follows \begin{equation}
		\hat{f} = \begin{pmatrix}
			- \frac{3 x^3}{2} \\
			-\frac{3yz}{2} \\
			3 z y^2
		\end{pmatrix}
	\end{equation} 
	Using the ansatz of Goriely-Hyde method, we get $ \mathbf{p} = (- \frac{1}{2}, - \frac{1}{2}, -1 ) $ and using these we get \footnote{At this point we would like to highlight that complex entries for the following $\mathbf{a}$ values and those observed in (15) are completely consistent with the fact that the system we have considered in (11-13) consists of x,y and z which are real and positive. As mentioned in section II, complex entries for various $\mathbf{a}$ suggest that the singularities will be non-finite time in nature and hence these quantities taking up complex values is consistent with the analysis as shown in \cite{goriely2000necessary}. Similar case has been for various cosmological systems (for example, see \cite{odintsov2018dynamical,odintsov2019finite})} \begin{equation}
		\begin{aligned}
			a_{1} = \left(\frac{1}{\sqrt{3}} , \frac{i}{\sqrt{3}} , \frac{1}{3} \right) \\[10pt]
			a_{2} = \left(\frac{1}{\sqrt{3}} , -\frac{i}{\sqrt{3}} , \frac{1}{3} \right) \\[10pt]
			a_{3} = \left(-\frac{1}{\sqrt{3}} , \frac{i}{\sqrt{3}} , \frac{1}{3} \right) \\[10pt]
			a_{4} = \left(-\frac{1}{\sqrt{3}} , -\frac{i}{\sqrt{3}} , \frac{1}{3} \right)
		\end{aligned}
	\end{equation}
	The Kovalevskaya matrix in this case takes the form \begin{equation}
		R =  \begin{pmatrix}
			\frac{1}{2} (1 - 9 x^2 ) & 0 & 0 \\
			0 & \frac{1}{2} (1-3z) & - \frac{3y}{2} \\
			0 & 6zy & 1 + 3 y^2
		\end{pmatrix}
	\end{equation}
	We can now evaluate the eigenvalues of this matrix in any set of the dominant balances mentioned above, to get \begin{equation}
		r = (-1, \sqrt{\frac{3}{2}}, -\sqrt{\frac{3}{2}})
	\end{equation}
	We again see that in this case we will have non finite time local singularities. 
	\\
	\\
	We consider one more truncation of this system, which is written as \begin{equation}
		\hat{f} = \begin{pmatrix}
			- \frac{3 xz}{2} \\
			-\frac{3y^3}{2} \\
			-3 z x^2
		\end{pmatrix}
	\end{equation}
	Using the usual ansatz, we get $ \mathbf{p} = (-\frac{1}{2},-\frac{1}{2},-1) $ and using this we get \begin{equation}
		\begin{aligned}
			a_{1} = \left(\frac{1}{\sqrt{3}} , \sqrt{\frac{2}{3}} , \frac{2}{3} \right) \\[10pt]
			a_{2} = \left(\frac{1}{\sqrt{3}} , -\sqrt{\frac{2}{3}} , \frac{2}{3} \right) \\[10pt]
			a_{3} = \left(-\frac{1}{\sqrt{3}} , \sqrt{\frac{2}{3}} , \frac{2}{3} \right) \\[10pt]
			a_{4} = \left(-\frac{1}{\sqrt{3}} , -\sqrt{\frac{2}{3}} , \frac{2}{3} \right)
		\end{aligned}
	\end{equation}
	We note now that in this truncation, various values of the dominant balance contain only real entries and so finite time singularities can occur in this system. Furthermore, the Kovalevskaya matrix takes the form The Kovalevskaya matrix in this case takes the form \begin{equation}
		R =  \begin{pmatrix}
			\frac{1}{2} (1 - 3z ) & 0 & - \frac{3x}{2} \\
			0 & \frac{1}{2} (1 - 9 y^2) & 0 \\
			- 6 z x & 0 & 1 - 3 x^2
		\end{pmatrix}
	\end{equation}
	One can now evaluate the eigenvalues of this matrix, in any set of the dominant balances as the same set of eigenvalues eventually come in, to arrive at \begin{equation}
		r = (-1,1,1)
	\end{equation}
	Here we see that $r_{2} = -1 $ and hence the singularity formed in this case will not be general. But this truncation tells us that the dynamical system allows finite time singularities for a limited set of initial conditions.
	\\
	\\
	Till now, we have discussed the singularity structure in this dark energy scenario from the dynamical point of view. But just pointing out that singularities could exist for this system would not be enough from the physical point of view and hence, one needs to classify properly about what kind of singularities could occur in this model. One can classify various types of physical singularities for cosmology at some time $ t = t_{s} $, where $t_{s} $ is the time at which the singularities occurs, as follows \cite{Nojiri:2005sx,Fernandez-Jambrina:2010ngm} 
	\begin{itemize}
		\item Type I ("Big Rip") : In this case, the scale factor a, effective energy density $ \rho_{eff} $ and effective pressure density $ p_{eff} $ diverges.
		\item Type II ("Sudden/Quiescent singularity") : In this case, $ p_{eff} $ diverges and so does the derivatives of the scalar factor from the second derivative onwards. 
		\item Type III ("Big Freeze") : In this case, the derivative of the scale factor from the first derivative onwards diverges 
		\item Type IV ("Generalized sudden singularities ") : In this case, the derivative of the scale factor diverges from a derivative higher than the second.
	\end{itemize}
	In this classification, Type I singularities are strong singularities in the sense that they can distort finite objects while singularities of Type II, Type III, and Type IV are weak singularities as they cannot be considered as either the beginning or the end of the universe. There are other minor types of singularities like w or Type V singularities, but we will only be considering type I- IV singularities here. The most general form of the Hubble parameter for studying singularities in the above classified types is \cite{odintsov2019finite} \begin{equation}
		H(t) = f_{1} (t) + f_{2} (t) (t - t_{s})^{\alpha}
	\end{equation}
	where $ f_{1} (t) $ and $ f_{2} (t) $ are assumed to be nonzero regular functions at the time of the singularity (similar conditions holding true for their derivatives till the second order) and $ \alpha $ is a real number. It is not necessary that the Hubble parameter (34) is a solution to the field equations, but we will consider the case that it is indeed true and will look for the implications of this consideration on the singularity structure in view of our dynamical analysis. First we note that none of x, y or z as defined in (10) can ever be singular for any value of cosmic time. The singularities that one can have considering the Hubble parameter as defined in (34) is \begin{itemize}
		\item For $\alpha < -1$, a big rip singularity occurs
		\item For $-1 < \alpha < 0 $, a type III singularity occurs
		\item For $ 0 < \alpha < 1 $ , a type II singularity occurs
		\item For $ \alpha > 1 $ , a type IV singularity occurs  
	\end{itemize} 
	To proceed further, we need to express $ \dot{\phi} $ and $ V(\phi) $ in terms of the Hubble parameter. For simplicity, we will consider that the coupling constant $ \mu = 1 $ and $ \dot{\rho_{B}} = 0 $. Making these considerations, we can write  \begin{equation}
		-2 \dot{H} = \dot{\phi}^2 \left(1 + \frac{\rho}{\sigma}\right)
	\end{equation}
	One can then write \begin{equation}
		\dot{\phi}^2 = -2 \left[ \left(\sigma +  V +  \sigma \rho_{B} \right) + \sqrt{\left( \sigma +  V +  \sigma \rho_{B} \right)^{2} - 2 \dot{H}} \right]
	\end{equation} 
	Furthermore, one can now write $ V(\phi)$ in terms of the dark energy equation of state \footnote{Note that here we are only considering dark energy equation of state with no background contributions, hence here we will only consider scalar field contributions} as \begin{equation}
		V(\phi) = \frac{\dot{\phi}^{2}}{2} \frac{(1-w)}{(1+w)}
	\end{equation}
	Using (36) and doing some algebra, we can write the potential now as \begin{equation}
		V = \frac{2 b (1+k) + \sqrt{(2b(1+k))^{2} - 2 \dot{H} (k^2 - 1)}}{2 (k^2 -1 )}
	\end{equation}
	where $k = \frac{2w}{1-w} $ and $b = \sigma (1+ \rho_{B}) $ (note that both k and b will always be positive for a positive brane tension). Notice that V is now completely in terms of the Hubble parameter (for constant values of $\sigma$, w and $\rho_{B} $) and so one can use this form of V in (36) to find $\dot{\phi}$ in terms of the Hubble parameter as well \footnote{While the potential here has been shown to be completely dependent on the Hubble parameter, interested readers would be tempted to explore various physically motivated forms of the potential for investigating the singularity structure. So it is important to stress the point here that V does depend crucially on whatever values w, the Brane tension and the bulk energy density will take and different values of these parameters will correspond to different physical scenarios. But since we will be interested only on the notion of regularity of the variables x,y and z, the various values of these parameters do not make much of a change.}. One can then use the form of the Hubble parameter as discussed in (34) in these expressions for $\dot{\phi}$ and V. It is necessary to express these quantities in terms of $H(t) $ as now we can find out which type of singularities are possible in this scenario, in the view of the fact that x,y and z as described in (10) have to remain regular. We will not be writing out the full expressions here for these variables but will comment on what type of singularities, as classified before, can occur in this scenario. By studying the expressions for x,y and z, one thing that is immediately clear is that Type I singularities are very easily allowed to occur in this scenario \footnote{At this point it is important to note the fact that we have already set an ansatz for the Hubble parameter in (26) and the values of $\alpha$ as described before determines what singularities can occur and what cannot. We emphasize this because a reader might be tempted to think that a type I singularity cannot occur as in accordance to (27), $\dot{H}$ will not be positive but the requirement for type I to occur is for $\alpha < -1 $  to be allowed given that x,y and z remain regular, which is certainly possible here. Similar case was seen \cite{odintsov2019finite}.} and hence a strong singularity like the big rip can take place given this model for dark energy which is an interesting assessment for the future of the universe in this scenario. Speaking of weak singularities, Type IV and Type III can occur as well but a Type II singularity cannot occur as there are terms $ \sim (t-t_{s})^{\alpha - 1 } $ in all of these expressions, which will be singular for $ 0 < \alpha < 1 $ . Note that we have not set any constraints on the value of the brane tension $\sigma$ nor on the equation of state parameter w (allowing for both phantom and quintessence scenarios) for the case in which it does not diverge and hence our treatment here is very general.
	\\
	\\
	In passing, we discuss here in detail the difference between the singularities obtained from the Goriely-Hyde method and the ones we have obtained from the ansatz (26). The singularity structure we discussed as above by considering the ansatz for the Hubble factor was made on the basis of the idea that the variable x,y and z remain regular for cosmic time, keeping in mind that the Hubble factor in denominator of all these variables can never vanish \footnote{So intrinsically we are ignoring Big bang/ Grand bang type singularities where the Hubble factor vanishes. }. The dynamical systems analysis through the Goriely-Hyde method shows us that both finite and infinite time singularities can indeed exist for a limited set of initial conditions for the variables x,y and z (10) but the question one could ask is, what would singularities in either of these variables mean physically for non vanishing  Hubble parameter ? Starting with z, if this variable becomes singular in finite or infinite time then this would mean that the energy density diverges and this could mean towards a type I singularity. Diverging energy density can also point towards a type II singularity if the considered equation of state is inhomogeneous \cite{Nojiri:2005sr}, but we're not considering inhomogeneous forms of the equation of state here. But in order for the energy density to be positive, z is constrained as $0\leq z \leq 1 $ with regards to the dynamical system (11-13). Hence z cannot become singular at any point in this scenario. Talking about x and y, there is no such constraint on x and y but if either of these values diverge then the it would mean that z is will also diverge and so if one of these variables diverge we need the other variable to diverge as well to cancel the singularities. This would mean that one of the variables will have to diverge towards -ve $\infty$ while one towards +ve $\infty$ and as y cannot be negative, y can only diverge positively and x can only diverge negatively. Hence even though the Goriely-Hyde procedure is very helpful in the sense that it allows us to study singularities in otherwise very complicated systems, there is a significant difference between the singularities obtained from the purely mathematical point of view and the ones we have pointed out from the physically motivated Hubble ansatz and it is important to pursue singularity from both ways to grip the limits of use of the mathematical technique from the physical view. It is also worth mentioning that in certain cases, quantum gravity based effects can also alleviate singularity formation in various cosmological settings \cite{Nojiri:2004ip,Nojiri:2005sx,Elizalde:2004mq}. In particular in \cite{Nojiri:2004ip} it was shown that conformal anomaly effects can be helpful for singularity removal even in an RS II Braneworld paradigm, so investigating whether quantum gravity effects can help in singularity removal in our considered scenario can be a very worthwhile endeavour and it could likely be the case although we will not be pursuing it in detail here. We would also like to point out that previous work on 5-d RS-II Braneworld models \cite{antoniadis2010brane,antoniadis2013brane} discussed how Type I and Type II singularities can occur in such scenarios (note that these works were not concerned with any dark energy models ) while here we have shown a different singularity structure for dark energy models in the RS-II paradigm. It is also worth noting that in \cite{Bamba:2008ut,Bamba:2010wfw} it was explicitly shown that $R^2$ terms may remove singularities (except of type IV one). For scalar fields with an exponential type potential, the same could be achieved . For fluids this is would correspond to a fluid (\cite{Bamba:2012cp}). As singularity removal is not in the focus of this paper, we have not endeavoured towards the same but the papers as discussed above show that it could be a viable possibility that the inclusion of R2 terms or conformal anomaly effects in different ways could help in removing some singularities in the model that we have considered too.              
	
	\section{Concluding remarks}
	In this paper we have discussed a regime of scalar field dark energy regime which is based in an RS-II Braneworld cosmology, focusing on the dynamical system that encapsulates the model. Such models of dark energy have been richly investigated in recent years and so we wanted to know the status quo of the occurrence of finite time singularities in such models. We employed a completely general approach here, where our treatment is not only limited to a particular type of potential but is valid for general potentials in this paradigm. We also considered implications from the swampland dS conjecture in this regard. We employed the Goriely-Hyde singularity analysis method to investigate finite time singularities in these models and found that they do admit finite time singularities for a limited set of initial conditions. We also found out that this result stands firm irrespective of whether or not one considers the dS conjecture. Finally, we then discussed about the physical nature of singularities that can occur in this scenario of dark energy. We showed that weak singularities like Type III and Type IV can occur in this scenario and also that a strong singularity like the Big Rip (Type I) can also occur in this case. We would also like to clarify that we have mainly considered FRW dynamics on the Brane and not in the bulk. We have given preference and more importance to the dynamics on the brane instead of at the bulk, for example, with our consideration that $w_{B}= 0$. We would refer the reader works like \cite{antoniadis2013brane,Deffayet:2000uy,Neves:2003eu} for more detailed discussions for the singularities and overall cosmological dynamics with emphasis on the bulk dynamics too. 
	\section*{Acknowledgements}
	The work by MK has been supported by the grant of the Russian Science Foundation No-18-12-00213-P https://rscf.ru/project/18-12-00213/ and performed in Southern Federal University. The authors would also like to thank Alain Goriely for his helpful comments on the singularity analysis method. We would also like to thank the reviewer of the paper for their insightful comments on the work.

	\bibliography{JSPJMJbrane.bib}

\providecommand{\noopsort}[1]{}\providecommand{\singleletter}[1]{#1}%
\begin{thebibliography}{10}

\bibitem{SupernovaSearchTeam:1998fmf}
Adam~G. Riess et~al.
\newblock {Observational evidence from supernovae for an accelerating universe
  and a cosmological constant}.
\newblock {\em Astron. J.}, 116:1009--1038, 1998.

\bibitem{Weinberg:1988cp}
Steven Weinberg.
\newblock {The Cosmological Constant Problem}.
\newblock {\em Rev. Mod. Phys.}, 61:1--23, 1989.

\bibitem{Lombriser:2019jia}
Lucas Lombriser.
\newblock {On the cosmological constant problem}.
\newblock {\em Phys. Lett. B}, 797:134804, 2019.

\bibitem{Copeland:2006wr}
Edmund~J. Copeland, M.~Sami, and Shinji Tsujikawa.
\newblock {Dynamics of dark energy}.
\newblock {\em Int. J. Mod. Phys. D}, 15:1753--1936, 2006.

\bibitem{Padmanabhan:2002ji}
T.~Padmanabhan.
\newblock {Cosmological constant: The Weight of the vacuum}.
\newblock {\em Phys. Rept.}, 380:235--320, 2003.

\bibitem{Capozziello:2011et}
Salvatore Capozziello and Mariafelicia De~Laurentis.
\newblock {Extended Theories of Gravity}.
\newblock {\em Phys. Rept.}, 509:167--321, 2011.

\bibitem{Nojiri:2010wj}
Shin'ichi Nojiri and Sergei~D. Odintsov.
\newblock {Unified cosmic history in modified gravity: from F(R) theory to
  Lorentz non-invariant models}.
\newblock {\em Phys. Rept.}, 505:59--144, 2011.

\bibitem{Nojiri:2017ncd}
S.~Nojiri, S.~D. Odintsov, and V.~K. Oikonomou.
\newblock {Modified Gravity Theories on a Nutshell: Inflation, Bounce and
  Late-time Evolution}.
\newblock {\em Phys. Rept.}, 692:1--104, 2017.

\bibitem{Zhang:2021ygh}
Zhen Zhang.
\newblock {Geometrization of light bending and its application to SdS$_{w}$
  spacetime}.
\newblock {\em Class. Quant. Grav.}, 39(1):015003, 2022.

\bibitem{Zlatev:1998tr}
Ivaylo Zlatev, Li-Min Wang, and Paul~J. Steinhardt.
\newblock {Quintessence, cosmic coincidence, and the cosmological constant}.
\newblock {\em Phys. Rev. Lett.}, 82:896--899, 1999.

\bibitem{Tsujikawa:2013fta}
Shinji Tsujikawa.
\newblock {Quintessence: A Review}.
\newblock {\em Class. Quant. Grav.}, 30:214003, 2013.

\bibitem{Faraoni:2000wk}
Valerio Faraoni.
\newblock {Inflation and quintessence with nonminimal coupling}.
\newblock {\em Phys. Rev. D}, 62:023504, 2000.

\bibitem{Gasperini:2001pc}
M.~Gasperini, F.~Piazza, and G.~Veneziano.
\newblock {Quintessence as a runaway dilaton}.
\newblock {\em Phys. Rev. D}, 65:023508, 2002.

\bibitem{Capozziello:2003tk}
Salvatore Capozziello, Sante Carloni, and Antonio Troisi.
\newblock {Quintessence without scalar fields}.
\newblock {\em Recent Res. Dev. Astron. Astrophys.}, 1:625, 2003.

\bibitem{Capozziello:2002rd}
Salvatore Capozziello.
\newblock {Curvature quintessence}.
\newblock {\em Int. J. Mod. Phys. D}, 11:483--492, 2002.

\bibitem{Carroll:1998zi}
Sean~M. Carroll.
\newblock {Quintessence and the rest of the world}.
\newblock {\em Phys. Rev. Lett.}, 81:3067--3070, 1998.

\bibitem{Caldwell:2005tm}
R.~R. Caldwell and Eric~V. Linder.
\newblock {The Limits of quintessence}.
\newblock {\em Phys. Rev. Lett.}, 95:141301, 2005.

\bibitem{Han:2018yrk}
Chengcheng Han, Shi Pi, and Misao Sasaki.
\newblock {Quintessence Saves Higgs Instability}.
\newblock {\em Phys. Lett. B}, 791:314--318, 2019.

\bibitem{Astashenok:2012kb}
Artyom~V. Astashenok, Shin'ichi Nojiri, Sergei~D. Odintsov, and Robert~J.
  Scherrer.
\newblock {Scalar dark energy models mimicking $\Lambda$CDM with arbitrary
  future evolution}.
\newblock {\em Phys. Lett. B}, 713:145--153, 2012.

\bibitem{Shahalam:2015sja}
M.~Shahalam, S.~D. Pathak, M.~M. Verma, M.~Yu. Khlopov, and R.~Myrzakulov.
\newblock {Dynamics of interacting quintessence}.
\newblock {\em Eur. Phys. J. C}, 75(8):395, 2015.

\bibitem{Colgain:2019joh}
Eoin~\'O. Colg\'ain and Hossein Yavartanoo.
\newblock {Testing the Swampland: $H_0$ tension}.
\newblock {\em Phys. Lett. B}, 797:134907, 2019.

\bibitem{Banerjee:2020xcn}
Aritra Banerjee, Haiying Cai, Lavinia Heisenberg, Eoin~\'O. Colg\'ain, M.~M.
  Sheikh-Jabbari, and Tao Yang.
\newblock {Hubble sinks in the low-redshift swampland}.
\newblock {\em Phys. Rev. D}, 103(8):L081305, 2021.

\bibitem{DiValentino:2021izs}
Eleonora Di~Valentino, Olga Mena, Supriya Pan, Luca Visinelli, Weiqiang Yang,
  Alessandro Melchiorri, David~F. Mota, Adam~G. Riess, and Joseph Silk.
\newblock {In the realm of the Hubble tension\textemdash{}a review of
  solutions}.
\newblock {\em Class. Quant. Grav.}, 38(15):153001, 2021.

\bibitem{riess2021comprehensive}
Adam~G Riess, Wenlong Yuan, Lucas~M Macri, Dan Scolnic, Dillon Brout, Stefano
  Casertano, David~O Jones, Yukei Murakami, Louise Breuval, Thomas~G Brink,
  et~al.
\newblock A comprehensive measurement of the local value of the hubble constant
  with 1 km/s/mpc uncertainty from the hubble space telescope and the sh0es
  team.
\newblock {\em arXiv preprint arXiv:2112.04510}, 2021.

\bibitem{Randall:1999ee}
Lisa Randall and Raman Sundrum.
\newblock {A Large mass hierarchy from a small extra dimension}.
\newblock {\em Phys. Rev. Lett.}, 83:3370--3373, 1999.

\bibitem{Randall:1999vf}
Lisa Randall and Raman Sundrum.
\newblock {An Alternative to compactification}.
\newblock {\em Phys. Rev. Lett.}, 83:4690--4693, 1999.

\bibitem{Deffayet:2000uy}
Cedric Deffayet.
\newblock {Cosmology on a brane in Minkowski bulk}.
\newblock {\em Phys. Lett. B}, 502:199--208, 2001.

\bibitem{Dvali:2000hr}
G.~R. Dvali, Gregory Gabadadze, and Massimo Porrati.
\newblock {4-D gravity on a brane in 5-D Minkowski space}.
\newblock {\em Phys. Lett. B}, 485:208--214, 2000.

\bibitem{Sahni:2002dx}
Varun Sahni and Yuri Shtanov.
\newblock {Brane world models of dark energy}.
\newblock {\em JCAP}, 11:014, 2003.

\bibitem{Sami:2004xk}
M.~Sami and V.~Sahni.
\newblock {Quintessential inflation on the brane and the relic gravity wave
  background}.
\newblock {\em Phys. Rev. D}, 70:083513, 2004.

\bibitem{Bento:2008yx}
M.~C. Bento, R.~Gonzalez Felipe, and N.~M.~C. Santos.
\newblock {Brane assisted quintessential inflation with transient
  acceleration}.
\newblock {\em Phys. Rev. D}, 77:123512, 2008.

\bibitem{Gonzalez:2008wa}
Tame Gonzalez, Tonatiuh Matos, Israel Quiros, and Alberto Vazquez-Gonzalez.
\newblock {Self-interacting Scalar Field Trapped in a Randall-Sundrum
  Braneworld: The Dynamical Systems Perspective}.
\newblock {\em Phys. Lett. B}, 676:161--167, 2009.

\bibitem{Leyva:2009zz}
Yoelsy Leyva, Dania Gonzalez, Tame Gonzalez, Tonatiuh Matos, and Israel Quiros.
\newblock {Dynamics of a self-interacting scalar field trapped in the
  braneworld for a wide variety of self-interaction potentials}.
\newblock {\em Phys. Rev. D}, 80:044026, 2009.

\bibitem{Escobar:2011cz}
Dagoberto Escobar, Carlos~R. Fadragas, Genly Leon, and Yoelsy Leyva.
\newblock {Phase space analysis of quintessence fields trapped in a
  Randall-Sundrum Braneworld: a refined study}.
\newblock {\em Class. Quant. Grav.}, 29:175005, 2012.

\bibitem{Nojiri:2004ip}
Shin'ichi Nojiri and Sergei~D. Odintsov.
\newblock {Quantum escape of sudden future singularity}.
\newblock {\em Phys. Lett. B}, 595:1--8, 2004.

\bibitem{Nojiri:2005sr}
Shin'ichi Nojiri and Sergei~D. Odintsov.
\newblock {Inhomogeneous equation of state of the universe: Phantom era, future
  singularity and crossing the phantom barrier}.
\newblock {\em Phys. Rev. D}, 72:023003, 2005.

\bibitem{Nojiri:2005sx}
Shin'ichi Nojiri, Sergei~D. Odintsov, and Shinji Tsujikawa.
\newblock {Properties of singularities in (phantom) dark energy universe}.
\newblock {\em Phys. Rev. D}, 71:063004, 2005.

\bibitem{Bamba:2008ut}
Kazuharu Bamba, Shin'ichi Nojiri, and Sergei~D. Odintsov.
\newblock {The Universe future in modified gravity theories: Approaching the
  finite-time future singularity}.
\newblock {\em JCAP}, 10:045, 2008.

\bibitem{Bamba:2010wfw}
Kazuharu Bamba, Sergei~D. Odintsov, Lorenzo Sebastiani, and Sergio Zerbini.
\newblock {Finite-time future singularities in modified Gauss-Bonnet and F(R,G)
  gravity and singularity avoidance}.
\newblock {\em Eur. Phys. J. C}, 67:295--310, 2010.

\bibitem{Nojiri:2008fk}
Shin'ichi Nojiri and Sergei~D. Odintsov.
\newblock {The Future evolution and finite-time singularities in F(R)-gravity
  unifying the inflation and cosmic acceleration}.
\newblock {\em Phys. Rev. D}, 78:046006, 2008.

\bibitem{odintsov2022did}
S.~D. Odintsov and V.~K. Oikonomou.
\newblock Did the universe experienced a pressure non-crushing type
  cosmological singularity in the recent past?, 2022.

\bibitem{goriely2000necessary}
Alain Goriely and Craig Hyde.
\newblock Necessary and sufficient conditions for finite time singularities in
  ordinary differential equations.
\newblock {\em Journal of Differential equations}, 161(2):422--448, 2000.

\bibitem{barrow2004more}
John~D Barrow.
\newblock More general sudden singularities.
\newblock {\em Classical and Quantum Gravity}, 21(23):5619, 2004.

\bibitem{cotsakis2007dominant}
Spiros Cotsakis and John~D Barrow.
\newblock The dominant balance at cosmological singularities.
\newblock In {\em Journal of Physics: Conference Series}, volume~68, page
  012004. IOP Publishing, 2007.

\bibitem{cotsakis2007asymptotics}
Spiros Cotsakis and Antonios Tsokaros.
\newblock Asymptotics of flat, radiation universes in quadratic gravity.
\newblock {\em Physics Letters B}, 651(5-6):341--344, 2007.

\bibitem{antoniadis2010brane}
Ignatios Antoniadis, Spiros Cotsakis, and Ifigeneia Klaoudatou.
\newblock Brane singularities and their avoidance.
\newblock {\em Classical and Quantum Gravity}, 27(23):235018, 2010.

\bibitem{antoniadis2013brane}
Ignatios Antoniadis, Spiros Cotsakis, and Ifigeneia Klaoudatou.
\newblock Brane singularities with mixtures in the bulk.
\newblock {\em Fortschritte der Physik}, 61(1):20--49, 2013.

\bibitem{antoniadis2014enveloping}
Ignatios Antoniadis, Spiros Cotsakis, and Ifigeneia Klaoudatou.
\newblock Enveloping branes and brane-world singularities.
\newblock {\em The European Physical Journal C}, 74(12):1--10, 2014.

\bibitem{odintsov2018dynamical}
Sergei~D Odintsov and Vasilis~K Oikonomou.
\newblock Dynamical systems perspective of cosmological finite-time
  singularities in f (r) gravity and interacting multifluid cosmology.
\newblock {\em Physical Review D}, 98(2):024013, 2018.

\bibitem{odintsov2019finite}
SD~Odintsov and VK~Oikonomou.
\newblock Finite-time singularities in swampland-related dark-energy models.
\newblock {\em EPL (Europhysics Letters)}, 126(2):20002, 2019.

\bibitem{vafa2005string}
Cumrun Vafa.
\newblock The string landscape and the swampland.
\newblock {\em arXiv preprint hep-th/0509212}, 2005.

\bibitem{Ooguri:2016pdq}
Hirosi Ooguri and Cumrun Vafa.
\newblock {Non-supersymmetric AdS and the Swampland}.
\newblock {\em Adv. Theor. Math. Phys.}, 21:1787--1801, 2017.

\bibitem{Agrawal:2018own}
Prateek Agrawal, Georges Obied, Paul~J. Steinhardt, and Cumrun Vafa.
\newblock {On the Cosmological Implications of the String Swampland}.
\newblock {\em Phys. Lett. B}, 784:271--276, 2018.

\bibitem{Chiang:2018lqx}
Chien-I Chiang, Jacob~M. Leedom, and Hitoshi Murayama.
\newblock {What does inflation say about dark energy given the swampland
  conjectures?}
\newblock {\em Phys. Rev. D}, 100(4):043505, 2019.

\bibitem{Trivedi:2020xlh}
Oem Trivedi.
\newblock {Implications of single field inflation in general cosmological
  scenarios on the nature of dark energy given the swampland conjectures}.
\newblock {\em Int. J. Geom. Meth. Mod. Phys.}, 18(14):2150231, 2021.

\bibitem{Dutta:2015jaq}
Jibitesh Dutta and H.~Zonunmawia.
\newblock {Complete cosmic scenario in the Randall-Sundrum braneworld from the
  dynamical systems perspective}.
\newblock {\em Eur. Phys. J. Plus}, 130(11):221, 2015.

\bibitem{Ravanpak:2020ewi}
Arvin Ravanpak and Golnaz~Farpour Fadakar.
\newblock {Dynamical System Analysis of Randall-Sundrum Model with Tachyon
  Field on the Brane}.
\newblock {\em Phys. Rev. D}, 101(10):103525, 2020.

\bibitem{Maartens:2010ar}
Roy Maartens and Kazuya Koyama.
\newblock {Brane-World Gravity}.
\newblock {\em Living Rev. Rel.}, 13:5, 2010.

\bibitem{Bouhmadi-Lopez:2002vho}
Mariam Bouhmadi-Lopez, Pedro~F. Gonzalez-Diaz, and Alexander Zhuk.
\newblock {On New gravitational instantons describing creation of brane
  worlds}.
\newblock {\em Class. Quant. Grav.}, 19:4863--4876, 2002.

\bibitem{Neves:2003eu}
Rui Neves and Cenalo Vaz.
\newblock {Brane world dynamics and conformal bulk fields}.
\newblock {\em Phys. Rev. D}, 68:024007, 2003.

\bibitem{Kanno:2007wj}
Sugumi Kanno, David Langlois, Misao Sasaki, and Jiro Soda.
\newblock {Kaluza-Klein braneworld cosmology with static internal dimensions}.
\newblock {\em Prog. Theor. Phys.}, 118:701--713, 2007.

\bibitem{Banerjee:2011wk}
Narayan Banerjee, Sayantani Lahiri, and Soumitra SenGupta.
\newblock {Cosmology in multiply warped braneworld scenario}.
\newblock {\em Int. J. Mod. Phys. A}, 29:1450069, 2014.

\bibitem{Rani:2021hvh}
Shamaila Rani and Nadeem Azhar.
\newblock {Braneworld Inspires Cosmological Implications of Barrow Holographic
  Dark Energy}.
\newblock {\em Universe}, 7(8):268, 2021.

\bibitem{Fernandez-Jambrina:2010ngm}
L.~Fernandez-Jambrina.
\newblock {$w$-cosmological singularities}.
\newblock {\em Phys. Rev. D}, 82:124004, 2010.

\bibitem{Elizalde:2004mq}
Emilio Elizalde, Shin'ichi Nojiri, and Sergei~D. Odintsov.
\newblock {Late-time cosmology in (phantom) scalar-tensor theory: Dark energy
  and the cosmic speed-up}.
\newblock {\em Phys. Rev. D}, 70:043539, 2004.

\bibitem{Bamba:2012cp}
Kazuharu Bamba, Salvatore Capozziello, Shin'ichi Nojiri, and Sergei~D.
  Odintsov.
\newblock {Dark energy cosmology: the equivalent description via different
  theoretical models and cosmography tests}.
\newblock {\em Astrophys. Space Sci.}, 342:155--228, 2012.

\end{thebibliography}
	
	\bibliographystyle{unsrt}
	
\end{document}